\documentclass[%
 reprint,
 amsmath,amssymb,
 aps,
 prx,
]{revtex4-2}

\usepackage{graphicx}
\usepackage{dcolumn}
\usepackage{svg}
\usepackage{bm}

\begin{document}

\preprint{APS/123-QED}

\title{Tube geometry controls protein cluster conformation and stability on the endoplasmic reticulum surface}


\author{Liam T. Kischuck}
\author{Aidan I. Brown}%
\email{aidan.brown@torontomu.ca}

\affiliation{Department of Physics, Toronto Metropolitan University, Toronto, Ontario, M5B 2K3, Canada}

\date{\today}

\begin{abstract}
 \label{abstract.tex}

The endoplasmic reticulum (ER), a cellular organelle that forms a cell-spanning network of tubes and sheets, is an important location of protein synthesis and folding. When the ER experiences sustained unfolded protein stress, IRE1 proteins embedded in the ER membrane activate and assemble into clusters as part of the unfolded protein response (UPR). We use kinetic Monte Carlo simulations to explore IRE1 clustering dynamics on the surface of ER tubes. While initially growing clusters are approximately round, once a cluster is sufficiently large a shorter interface length can be achieved by `wrapping' around the ER tube. A wrapped cluster can grow without further interface length increases. Relative to wide tubes, narrower tubes enable cluster wrapping at smaller cluster sizes. Our simulations show that wrapped clusters on narrower tubes grow more rapidly, evaporate more slowly, and require a lower protein concentration to grow compared to equal-area round clusters on wider tubes. These results suggest that cluster wrapping, facilitated by narrower tubes, could be an important factor in the growth and stability of IRE1 clusters and thus impact the persistence of the UPR, connecting geometry to signaling behavior. This work is consistent with recent experimental observations of IRE1 clusters wrapped around narrow tubes in the ER network. 
\end{abstract}

\maketitle


\section{Introduction}

Protein clustering can be important to the health and function of a living cell. While many instances of protein clustering in cell biology occur in three-dimensional volumes~\cite{berry2018physical}, protein clusters can also form on the two-dimensional surface of the cell membrane~\cite{hartman2011signaling}, such as those of nephrin~\cite{banjade2014phase} and GPCR proteins~\cite{panetta2008physiological}; and on various intracellular membranes, such as DRP1 and MAVS proteins on mitochondria~\cite{frohlich2013structural,ugarte2014dynamin,hou2011mavs} and NBR1 proteins on peroxisomes~\cite{brown2015cluster,brown2017model}. Coarsening dynamics, including cluster coalescence and exchange of material between clusters, often describe cluster evolution following formation~\cite{brown2015cluster,berry2018physical}.


Clustering of the protein IRE1 on the endoplasmic reticulum (ER) membrane activates the unfolded protein response (UPR) that maintains protein folding homeostasis inside the ER~\cite{li2010mammalian, belyy2020quantitative,brown2021design}. The ER is an organelle composed of a cell-spanning network of tubes and sheets. A substantial fraction of proteins produced by the cell transit through the ER, and many are sent from ER exit sites to the Golgi for further processing~\cite{westrate_form_2015}. While chaperones and other ER factors assist nascent proteins in folding into functional conformations and the ER contains quality control pathways to remove misfolded or excess unfolded proteins~\cite{brown2021design}, unfolded proteins can accumulate in the ER and impede the function of the ER and the cell~\cite{hetz2020mechanisms}. Accumulation of unfolded proteins in the ER network, known as ER stress, triggers the UPR, which induces changes to mitigate the unfolded protein stress.

There are multiple UPR signaling pathways, and the pathway mediated by IRE1 proteins is the most ancient and widely conserved, shared across yeast, plants, and mammals~\cite{zhang2016divergence}. IRE1 is activated by removal of an attached BiP chaperone by an unfolded protein, followed by autophosphorylation that enables the formation of IRE1 dimers and higher order oligomers~\cite{hetz2012unfolded,korennykh2009unfolded}. IRE1 oligomers splice mRNA which then traffics into the nucleus to modify gene expression~\cite{hetz2012unfolded} and mRNA turnover \cite{hetz2020mechanisms,le2021decoding}. Under conditions of prolonged activity, IRE1 can play a role in the activation of apoptosis (programmed cell death)~\cite{hetz2018unfolded}.
The IRE1 activation from ER stress causes IRE1 clustering on the membrane of ER tubes \cite{kimata2007two,korennykh2009unfolded,ricci2019clustering,belyy2020quantitative}. This clustering behavior is thought to be essential to the UPR signaling of IRE1~\cite{li2010mammalian,korennykh2009unfolded}.
The proteins at the cluster periphery diffuse in and out of the cluster, and evaporating clusters appear to break apart at their edges \cite{belyy2020quantitative}.

With IRE1 present on the ER membrane in relatively low concentrations ($\sim$ 1/$\mu$m$^2$ in both mammals~\cite{belyy2020quantitative} and yeast~\cite{aragon2009messenger,west20113d,brown2022mitochondrial}), the factors that control diffusive encounters between IRE1 proteins and clusters are expected to impact IRE1 clustering behavior. While mean search times can be calculated on spatial networks such as the ER~\cite{brown2020impact,scott2021diffusive}, the diffusive search by activated IRE1 proteins for other IRE1 monomers and clusters is expected to exhibit large variation~\cite{grebenkov2018strong}. IRE1 clusters form in yeast in $\sim$10 minutes~\cite{aragon2009messenger} and in mammals in $\sim$2 hours~\cite{belyy2020quantitative}. With IRE1 diffusivity of $D = 0.24\text{ }\mu\text{m}^2/\text{s}$~\cite{belyy2020quantitative}, diffusive search times for a target in the mammalian ER would be on the scale of an hour~\cite{brown2020impact}, consistent with diffusive search playing a role in determining cluster formation times. IRE1 clusters continue to evolve in both yeast and mammals over $\sim$10 hours~\cite{aragon2009messenger,cohen2017iron,belyy2020quantitative}. This slow decrease of cluster number~\cite{cohen2017iron,belyy2020quantitative} and increase in cluster size after cluster formation~\cite{belyy2020quantitative}, and large majority of IRE1 localized to clusters~\cite{aragon2009messenger} are consistent with control of cluster evolution by Ostwald ripening interactions between clusters.

IRE1 clusters are confined to the two-dimensional ER membrane surface.
It is expected that geometric confinement and search dimension will affect diffusive search and ~\cite{berg1977physics,benichou2010geometry,brown2020impact,condamin2007first,koslover2011theoretical} and phase separation behavior~\cite{krishnamachari1996gibbs,brown2015cluster,brown2017model,yao1993theory}, key aspects of clustering activity.
IRE1 clusters have been observed with diverse morphologies, including those that appear wrapped around ER tubes~\cite{belyy2020quantitative,tran2021stress} and localize to narrow ER tubes~\cite{tran2021stress}. 

Given the novel IRE1 clustering behavior under ER stress conditions, we use kinetic Monte Carlo simulations to explore how ER tube geometry impacts IRE1 clustering behavior. Because of the unique periodic geometry of a tube, sufficiently large clusters can `wrap around' a tube, and reduce the interface length compared to an equal-area round cluster. The focus of this work is to describe how cluster wrapping impacts cluster dynamics. This `wrapped' cluster morphology is directly impacted by tube geometry, as narrower tubes allow smaller clusters to become wrapped. We find that tube radius, by controlling the transition between wrapped and circular IRE1 clusters, impacts cluster growth rate and stability (evaporation rate and threshold between cluster growth and evaporation). This geometric effect on cluster dynamics is expected to be a significant factor for IRE1 cluster behavior and the persistence of  intracellular UPR signaling.




\section{Results}

\subsection{Transition between circular and wrapped IRE1 clusters} \label{closed.tex}

\begin{figure}[h!] \begin{center}
\includegraphics[width = 0.45\textwidth]{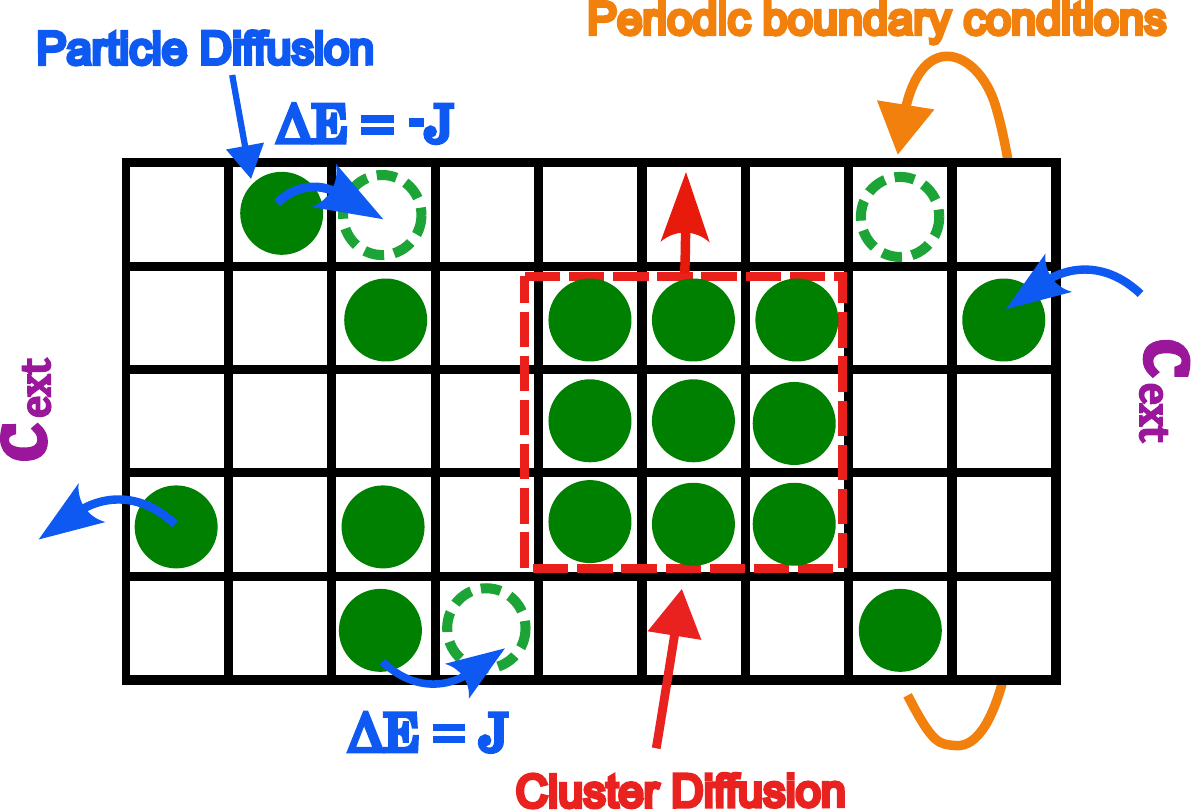}
\caption{Schematic diagram of the kinetic Monte Carlo simulation algorithm for IRE1 protein and cluster dynamics on an endoplasmic reticulum tube. Green circles are individual activated IRE1 proteins, which diffuse on a two-dimensional lattice as individual proteins and as clusters. Periodic boundary conditions (orange arrow) represent the tubular geometry. Connection to the rest of the ER network is represented by a constant external concentration $c_{\text{ext}}$, with proteins able to both enter and leave the tube section. IRE1 interactions are favored by nearest-neighbor interaction energy $J$ (see Eq.~\ref{eq:energy}).}
\label{fig:KMC}
\end{center}
\end{figure}

\begin{figure*}[] 
\includegraphics[width = 0.85\textwidth]{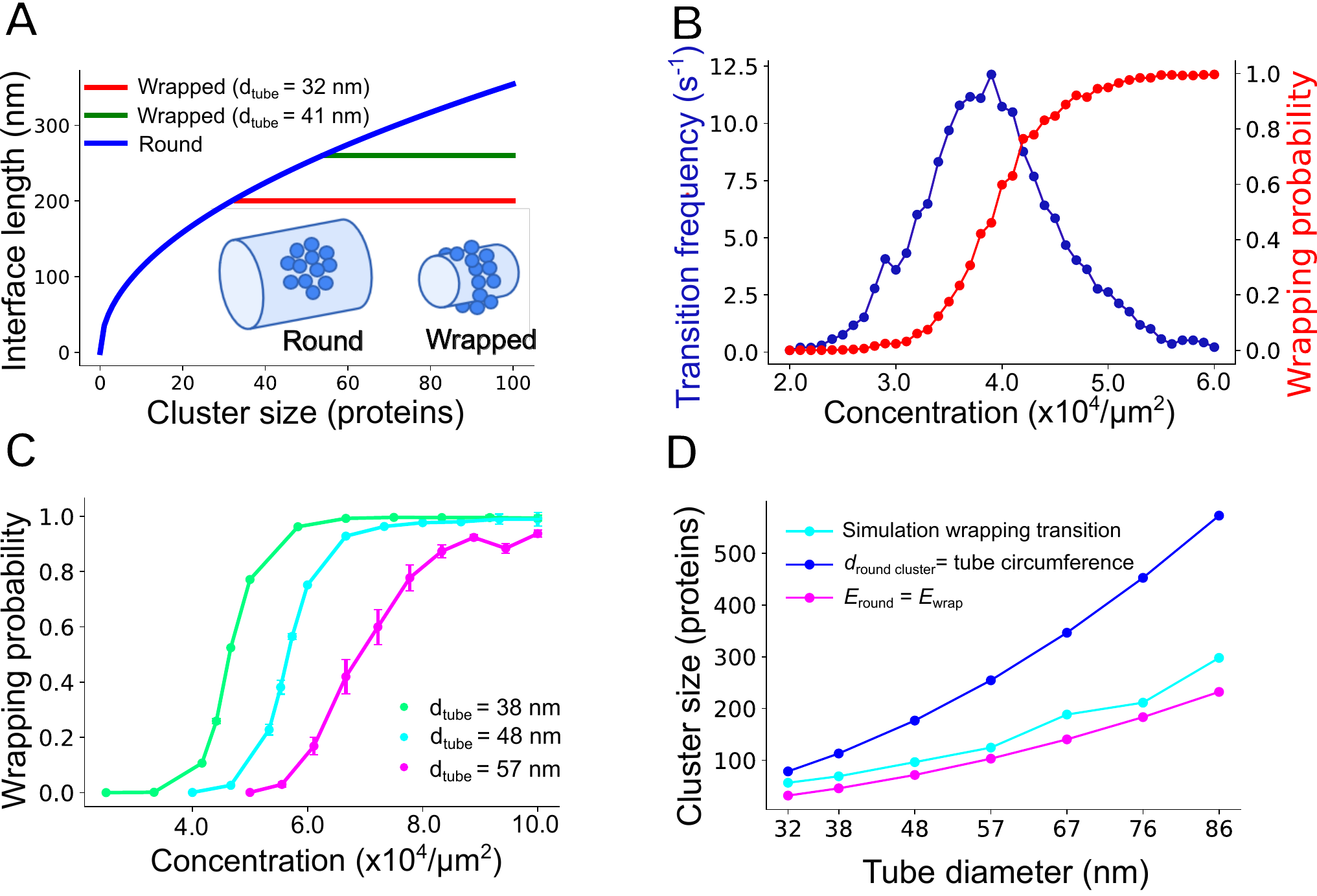}
\caption{Cluster conformation transitions.
(A) Schematics of round (left inset) and wrapped (right inset) cluster conformations. Cluster interface lengths for round (blue), and wrapped (red and green) clusters on tubes (diameters 32 nm and 41 nm).
(B) Frequency of transitions between round and wrapped clusters on a 32 nm diameter and 1 $\mu$m closed long tube, averaged over 24 runs; and corresponding probability of wrapped cluster conformation.
(C) Probability of cluster conformation (wrapped or round) on a closed tube (no proteins enter or exit) for various protein concentrations, with 38, 48 and 57 nm tube diameter and 1 $\mu$m length. Simulations begin with all proteins in a round cluster at tube center. $J= 3\text{ }k_{\text{B}}T$, average of 24 runs. 
(D) Cluster size at which the interface energy of wrapped clusters becomes energetically favored compared to round clusters (magenta, Eq.~\ref{wrappingSize}), cluster size at which 50$\%$ of clusters are in a wrapped conformation in simulations from (C) (cyan), and cluster size at which round cluster diameter is equal to tube circumference (blue).}
\label{fig:ClosedSystem}
\end{figure*}


IRE1 protein dynamics on a two-dimensional lattice representing a single tube region of the ER are simulated using the kinetic Monte Carlo method~\cite{bortz1975new,andersen2019practical,cheimarios2021monte}. This approach is similar to previous lattice-gas models of protein clustering on cellular membranes~\cite{ryan2007clocking}. All IRE1 in this region are assumed activated, in line with the global IRE1 activation observed experimentally~\cite{belyy2020quantitative, korennykh2009unfolded} or with local IRE1 activation~\cite{acosta2018unfolded}. The simulation includes (see Fig.~\ref{fig:KMC}) diffusion of both individual IRE1 proteins and IRE1 clusters, with each lattice site permitted to be occupied by one protein or not occupied. Following experiments showing diffusivity in a membrane scaling with inverse particle radius~\cite{gambin2006lateral} and consistent with the Stokes-Einstein relation~\cite{luo2015kinetic}, cluster diffusivity is scaled as $N^{-1/2}$, where $N$ is the number of proteins in the cluster. The two-dimensional lattice is periodic in one direction to represent the tubular geometry. In the other direction, IRE1 that diffuses past either end of the two-dimensional lattice has left the tube region under consideration.
Proteins enter both tube ends proportional to an external concentration $c_{\text{ext}}$. Nearest-neighbor IRE1 proteins have a favorable interaction energy, with the energy $E$ of the system
\begin{equation}
\label{eq:energy}
E = -J\sum_{\langle ij\rangle} n_i n_j \ ,
\end{equation}
where $J>0$, $\langle ij\rangle$ representing the sum over nearest-neighbor lattice sites, and $n_i\in[0,1]$ is the occupation of lattice site $i$. Diffusive moves of individual IRE1 proteins to unoccupied sites occur at rate $k = k_0\text{min}[1,\exp(-\Delta E/(k_{\text{B}}T))]$, with $k_0$ the rate in the absence of other proteins, and $\Delta E$ the change in $E$ (Eq.~\ref{eq:energy}) from the move, following the Metropolis criterion~\cite{metropolis1949monte}. Clusters can additionally take collective diffusive steps to unoccupied sites with $k_0$ reduced by a factor $\sqrt{N}$.

While IRE1 clusters that are small compared to the ER tube circumference are expected to be approximately circular in shape (`round'), experiments show that IRE1 proteins can form clusters that wrap around an ER tube~\cite{tran2021stress,belyy2020quantitative}. The inset of Fig.~\ref{fig:ClosedSystem}A shows a schematic of these two types of cluster shape.
The interface length (circumference) of a round cluster is  $L_{\text{round}} = 2\pi \sqrt{Na/\pi}$, where $a$ is the membrane area occupied by one IRE1 protein. For a cluster that is sufficiently large to wrap around the tube, the interface length of a wrapped cluster is $L_{\text{wrap}} = 4\pi r_{\text{tube}}$, where $r_{\text{tube}}$ is the ER tube radius. Note that this wrapped cluster interface length is independent of the number of proteins in the cluster, in contrast to the protein number-dependent interface length of the round cluster. These interface lengths are shown in Fig.~\ref{fig:ClosedSystem}A for a 32 nm and 41 nm tube radius as the cluster size changes. Depending on the radius of the ER tube, if the cluster can wrap around the tube, a wrapped cluster will have a shorter interface length.

Because cluster energy is from nearest-neighbor interactions (Eq.~\ref{eq:energy}), the cluster energy $E_{\text{cluster}} \approx -2JN +LJ/\sqrt{a}$ is approximately the bulk interactions expected if all proteins were surrounded by nearest neighbors ($-2JN$) plus the `missing' interactions from proteins at the interface rather than surrounded by other proteins in the bulk ($LJ/\sqrt{a}$). Thus the energy of a cluster of a given protein number $N$ is proportional to the interface length $L$, and a shorter-interface cluster of the same protein number will have a lower energy. Corresponding to cluster sizes for which the round cluster interface length will be longer than the wrapped cluster interface length, the round cluster energy will exceed the energy of a wrapped cluster for
\begin{equation}
    N > \frac{4 \pi r^2_{\text{tube}}}{a} \ .
    \label{wrappingSize}
\end{equation}

To explore transitions between round and wrapped clusters, we simulated IRE1 cluster dynamics on a closed tube with proteins unable to leave or enter (see Appendix for snapshots of a sample transition). While for a given cluster size and tube radius, either a wrapped or round cluster is energetically favored, there may not be a large enough energy difference for the cluster to indefinitely settle into either the wrapped or round conformation. The frequency of transitions between wrapped and round clusters is maximized at an intermediate protein concentration, with round clusters favored at lower concentrations and wrapped clusters at higher concentrations (Fig.~\ref{fig:ClosedSystem}B). The transition frequency peak indicates a phase transition between round and wrapped conformations. Figure~\ref{fig:ClosedSystem}C shows how the probability of finding a cluster in the the wrapped conformation increases with protein concentration (and thus cluster size), transitioning from a low wrapped probability to high over a relatively narrow range of protein concentrations. The transition concentration from round to wrapped clusters increases as the tube radius increases, as increased tube radii require larger clusters for a wrapped conformation to be favored (Eq.~\ref{wrappingSize}).


Figure \ref{fig:ClosedSystem}D plots the cluster size $N$ at which wrapped clusters become favored, described by Eq.~\ref{wrappingSize}, and compares to the cluster size for this transition observed in simulations. While the simulation transition from round to wrapped follows a similar trend to the prediction, the simulation transitions occur for slightly higher cluster sizes. This is expected due to fluctuations in cluster size from proteins joining/leaving, and from fluctuations in cluster conformation as a wrapped cluster temporarily unwraps. Clusters wrap around the tube for substantially smaller cluster sizes than would naturally wrap through growth of a round cluster, i.e., at smaller cluster size than for cluster diameter = tube circumference (Fig.~\ref{fig:ClosedSystem}D).
\subsection{Cluster growth depends on tube diameter} \label{growth.tex}


\begin{figure}[] 
\begin{center}
\includegraphics[width = 0.47\textwidth]{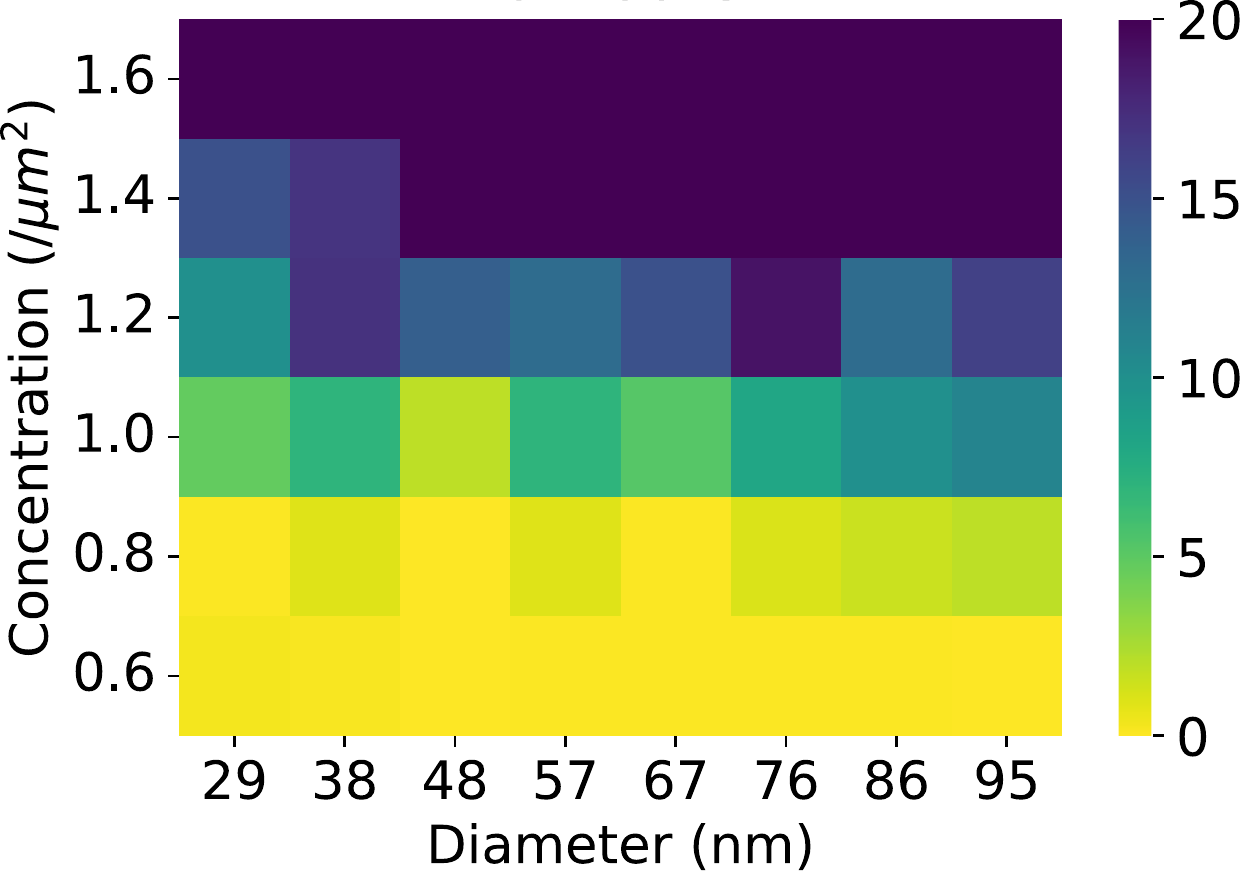}
\caption{Cluster formation with open boundary conditions. The IRE1 external (to the tube region under consideration) concentration $c_{\text{ext}}$ and the tube diameter are varied, with color map indicating mean number of proteins in a cluster 6 hours after initializing with one protein in the tube, averaged over 20 runs, tube length of 1 $\mu$m, and $J = 5.3\text{ }k_{\text{B}}T$. Cluster sizes over 20 are indicated as 20. Each tube initially contains a single IRE1 protein, as the maximum diameter and concentration sampled correspond to less than a single IRE1 protein in the tube region.}
\label{fig:OpenPD}
\end{center}

\end{figure}

We now move from exploring IRE1 cluster dynamics in a closed system with a pre-formed cluster to cluster formation and growth behavior. In these simulations the external IRE1 concentration is held fixed, with proteins able to enter and leave the tube region under consideration.
IRE1 clusters form through monomer dimerization followed by further growth and merging of small multimers.
At low external IRE1 concentrations, clusters do not form. Cluster formation occurs above a concentration threshold of approximately $1/\mu\text{m}^2$ independent of tube diameter for $J = 5.3\text{ }k_{\text{B}}T$, shown in  Fig.~\ref{fig:OpenPD} after a period of 6 hours. Similar results are found at earlier times, suggesting cluster formation at different tube diameters and concentrations is no longer changing with time (see Appendix). The independence of the cluster formation concentration threshold on tube diameter is attributed to the lack of control of the diameter on the formation and encounter kinetics of monomers and small multimers.

\begin{figure*}
\includegraphics[width = 0.85\textwidth]{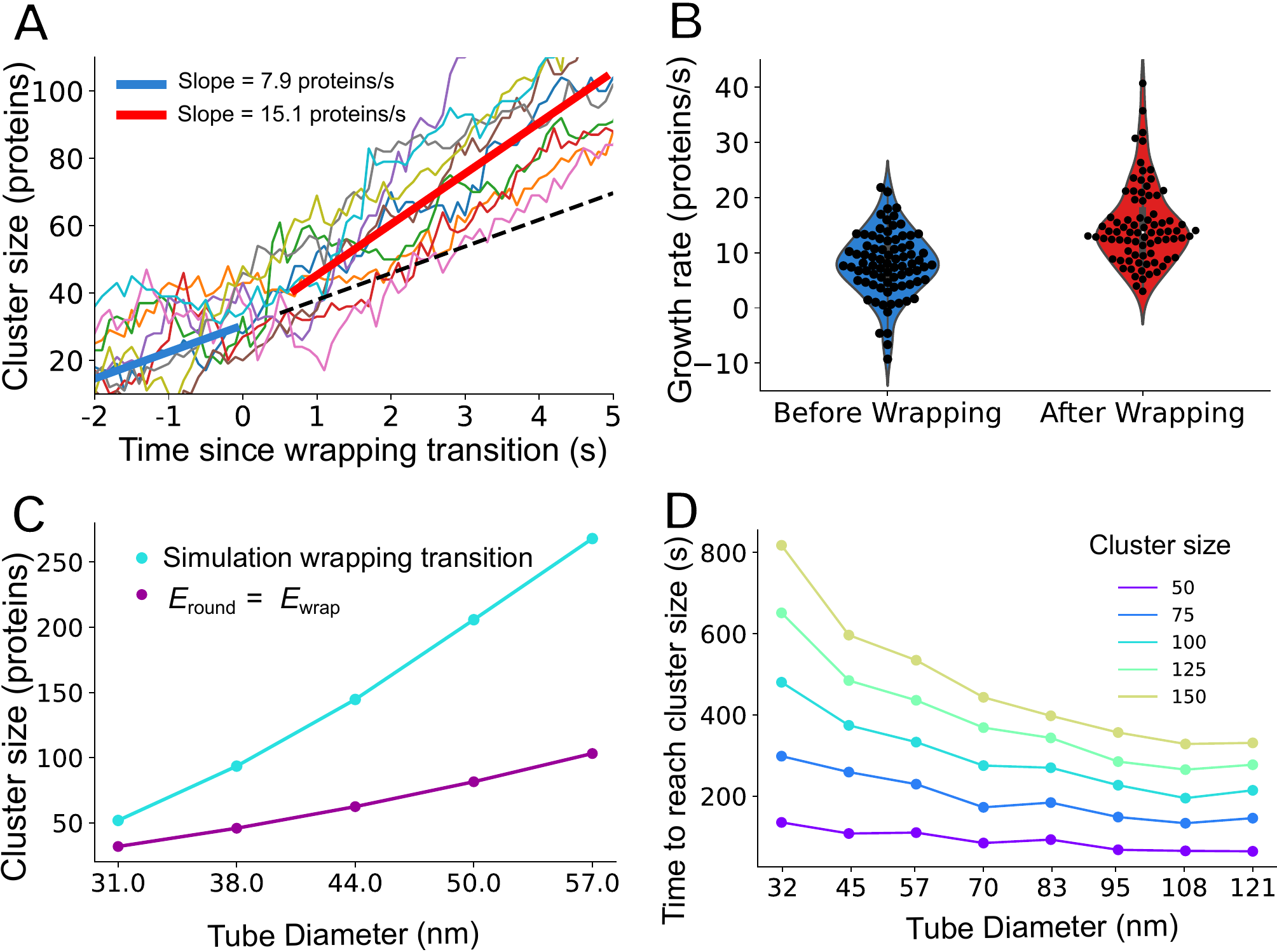}
\caption{Cluster conformation affects cluster growth rate.
(A) Individual cluster growth trajectories (thin lines) vs time, as well as linear fits of 80 cluster size trajectories before (thick blue) and after (thick red) round-to-wrapped transition, which is set at $t=0$. $J = 3\text{ }k_BT$, $c_{\text{ext}} = 34/\mu\text{m}^2$, 30 nm tube diameter, and tube length of $1\text{ }\mu\text{m}$.
(B) Violin plot of the cluster growth rates before and after the cluster wrapping transition, with same parameters as (A).
(C) Cluster size at wrapping transition vs tube diameter, with cluster size at which the interface energy of wrapped clusters becomes energetically favorable (purple, Eq.~\ref{wrappingSize}) and cluster size at which wrapping transition occurs during cluster growth simulations (cyan). Tube length 1 $\mu$m, $J=5.3\text{ }k_BT$, averaged over 30 samples.
(D) Mean time period for a cluster to grow to a specific size (see legend) vs tube diameter. $J=5.3\text{ }k_BT$, $c_{\text{ext}} = 1/\mu\text{m}^2$, tube length of $1\text{ }\mu\text{m}$, averaged over 30 samples with initial cluster size of 30 proteins.}
\label{fig:OpenSystem}
\end{figure*}

We next examine the growth of IRE1 clusters over time, in these same open tubes exposed to a constant external concentration. 
When the cluster is small, the cluster has a round conformation as it grows. Once sufficiently large, a cluster will fluctuate into a wrapped conformation. The average cluster growth rate can substantially increase following this transition from a round to a wrapped conformation (Fig.~\ref{fig:OpenSystem}A). The cluster growth rate distribution before the wrapping transition is distinct from the distribution after the transition (Fig.~\ref{fig:OpenSystem}B), with the likelihood of these growth distributions being drawn from the same underlying distribution approximately $10^{-7}$ according to the Kolmogorov-Smirnov test. While the choice of parameters in Fig.~\ref{fig:OpenSystem}A,B emphasize the cluster growth rate difference, other parameter choices yield a smaller but consistent growth advantage for clusters following the wrapping transition (see Appendix). The increased growth rate of wrapped clusters, compared to round clusters on the same tube, is attributed to an interface that no longer grows with size (no further increase in cluster escape), a flat interface compared to round clusters (reduces cluster escape), and the encounter of all proteins with a wrapped cluster, compared to a round cluster that a protein can diffuse past without encountering the cluster.

Figure~\ref{fig:OpenSystem}C shows the cluster size at which wrapped clusters become energetically favored (Eq.~\ref{wrappingSize}) and the cluster size at which the wrapping transition occurs in simulations with a constant external concentration. The wrapping transition occurs in the simulated growing clusters at a significantly larger cluster size than when the wrapped conformation becomes favorable compared to the round conformation (compare to Fig.~\ref{fig:ClosedSystem}D).
This widening of the gap for growing clusters (Fig.~\ref{fig:OpenSystem}C) compared to a closed system (Fig.~\ref{fig:ClosedSystem}D) is expected, as the fluctuation from a round to a wrapped conformation requires finite time, and if the cluster is continuing to grow the cluster size will be larger when the transition finally occurs.

We then investigated whether this increase in cluster growth rate after the wrapping transition leads to tube-radius dependent cluster growth rates.
Figure \ref{fig:OpenSystem}D shows that clusters grow more rapidly on wider tubes compared to smaller tubes.  
With the external IRE1 concentration per membrane area held constant, wider tubes (with larger circumferences) will have a larger number of IRE1 enter per unit time, leading to faster cluster growth.

While clusters on a given tube diameter grow faster after wrapping (Fig.~\ref{fig:OpenSystem}A,B), clusters reach a certain size more quickly on wider tubes that disfavor the wrapping transition (Fig.~\ref{fig:OpenSystem}D). More proteins enter wider tubes because of their larger circumference exposed to the same constant area concentration of proteins as narrower tubes. The additional proteins provided to wide tubes allows clusters on wider tubes to more quickly reach a certain size, compared to narrow tubes with clusters that wrap at smaller cluster sizes. The time for a cluster to grow to a certain size incorporates delivery rates, which are larger for wider tubes; as well as cluster conformation, with enhances growth after the transition to a wrapped cluster, which occurs earlier in cluster growth on narrower tubes.


\subsection{Wrapped clusters decay more slowly than round clusters} \label{decay.tex}

\begin{figure*}
\includegraphics[width = \textwidth]{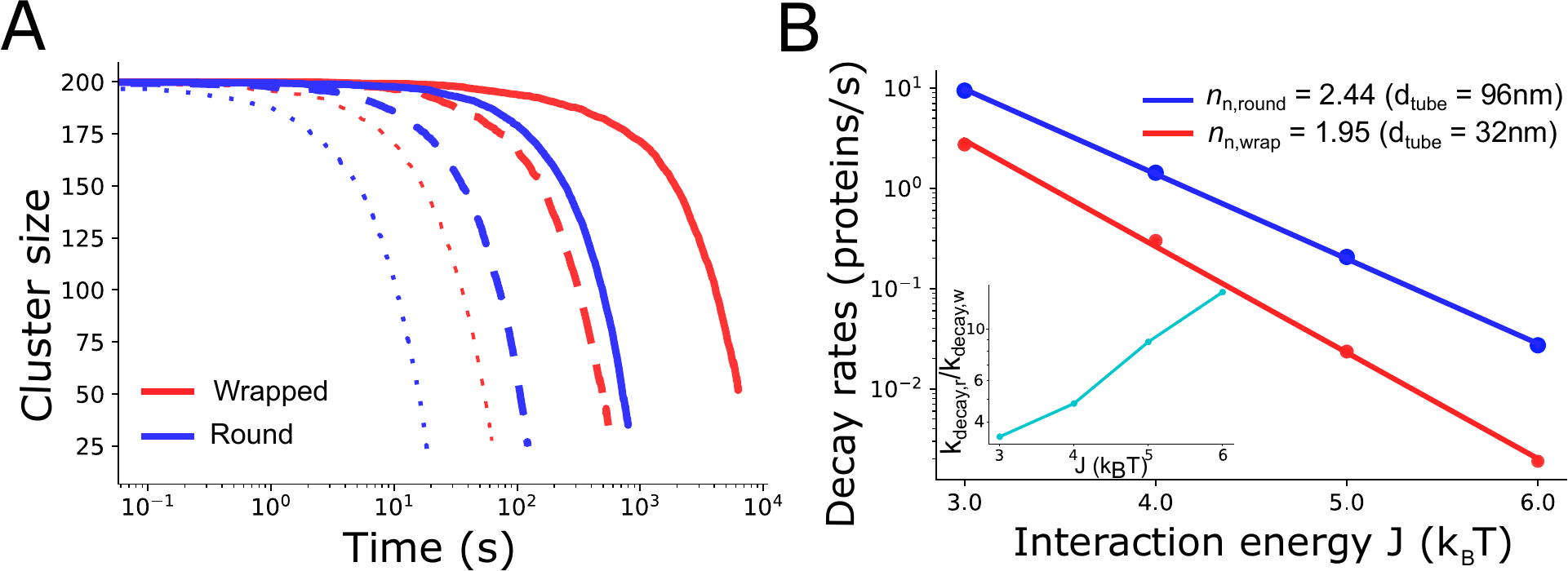}
\caption{Cluster decay.
(A) Mean cluster size vs time with external concentration $c_{\text{ext}} = 0$. Red lines are a wrapped cluster on a wide (96 nm diameter) tube and blue lines are a round cluster on a narrow (32 nm diameter) tube. $J = 5\text{ }k_{\text{B}}T$ (solid lines), $J = 4\text{ }k_{\text{B}}T$ (dashed), and $J=3\text{ }k_{\text{B}}T$ (dotted). Cluster size is mean over 30 samples, with clusters initially 200 proteins, 1 $\mu$m tube length.
(B) Cluster decay rates from decay curves in (A). Indicated $n_{\text{n,round}}$ and $n_{\text{n,wrap}}$ found by linear regression. Inset shows ratio of round cluster (wide tube) decay rate to wrapped cluster (narrow tube) decay rate, $k_{\text{decay,r}}/k_{\text{decay,w}}$.}

\label{fig:Decay}
\end{figure*}


We now move from exploring cluster growth to cluster decay by setting the external IRE1 concentration to zero. Therefore, during these cluster decay processes, IRE1 proteins can exit the tube region under consideration, but do not enter. 

Figure~\ref{fig:Decay}A shows cluster decay for round and wrapped clusters. All clusters initially have 200 proteins, with round clusters on a wider (96 nm diameter) tube and wrapped clusters on a narrower (32 nm) tube.
Wrapped clusters decay much more slowly, with 200 protein clusters in a wrapped conformation decaying to 50 proteins after approximately $9\times$ ($J = 5\text{ }k_{\text{B}}T$), $5\times$ ($J = 4\text{ }k_{\text{B}}T$), and $3.5\times$ ($J = 3\text{ }k_{\text{B}}T$) longer time periods compared to round clusters.
The cluster decay rates for wrapped and round clusters are compared across interaction energies in Fig.~\ref{fig:Decay}B. While wrapped clusters consistently decay more slowly than round clusters, the ratio between the wrapped and round cluster decay rates rises as the IRE1 interaction energy $J$ increases (Fig.~\ref{fig:Decay}B inset).

Proteins evaporating from a cluster increase in energy by $n_{\text{n}}J$, where $n_{\text{n}}$ is the typical number of neighbors of a protein on the cluster surface, such that the protein escape rate is proportional to $e^{-n_{\text{n}}J/(k_{\text{B}}T)}$. $n_{\text{n}}$ is estimated from the cluster decay (see Fig.~\ref{fig:Decay}B), with 200-protein wrapped clusters exhibiting $n_{\text{n,wrap}} = 2.44$ and 200-protein round clusters $n_{\text{n,round}} = 1.95$. This lower number of neighbors for round clusters aligns with the increased interface length and curvature for round clusters relative to the flatter wrapped cluster interface, which leads to increased protein escape rates from the surface of round clusters relative to wrapped clusters~\cite{krishnamachari1996gibbs}. The ratio of the decay rates is an exponential,
\begin{equation}
    \frac{e^{-n_{\text{n,round}}J/(k_{\text{B}}T)}}{e^{-n_{\text{n,wrap}}J/(k_{\text{B}}T)}} = e^{\Delta n J/(k_{\text{B}}T)} \ ,
\end{equation}
with $\Delta n = n_{\text{n,wrap}} - n_{\text{n,round}}$ as shown in the inset of Fig.~\ref{fig:Decay}B. By selecting the cluster conformation (wrapped or round), tube width controls the speed of cluster evaporation.




\subsection{Wrapped clusters have lower threshold protein concentrations between growth and decay} \label{hyst.tex}

\begin{figure*}
\includegraphics[width = \textwidth]{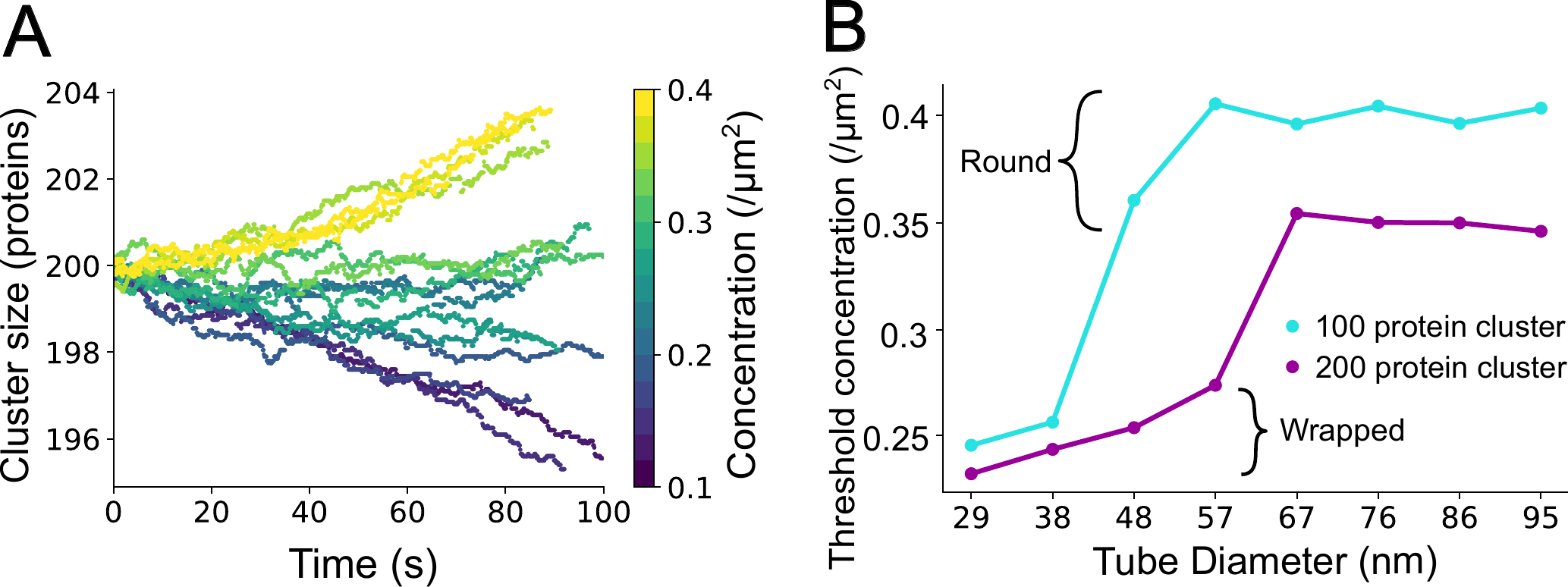}
\caption{Threshold concentration between cluster growth and decay.
(A) Mean trajectories for clusters in tube with different external concentrations $c_{\text{ext}}$ (each curve is a different $c_{\text{ext}}$). The lowest concentration is $0.1 / \mu \text{m}^2$, highest concentration is $0.4/\mu \text{m}^2$, with concentration intervals of approximately $0.021/\mu\text{m}^2$, with 15 concentrations shown. Each concentration averaged over 20 runs. Clusters begin with 200 proteins, and begin as round unless tube is too narrow, in which case the cluster starts wrapped. 
(B) Threshold concentration $c_0$ between cluster growth and decay as tube diameter is varied. Threshold concentration determined by linear fit of  cluster growth rates near zero cluster growth. For diameters less than the diameter at which the threshold concentration transitions (suddenly jumps), clusters are wrapped; for diameters greater than the transition, clusters are round. Tube length 1 $\mu$m, $J = 5.3 k_BT$ as in the phase diagram of Fig.~\ref{fig:OpenPD}.}
\label{fig:Threshold}
\end{figure*}



We determine the threshold external IRE1 concentration $c_{\text{ext}}$ at which simulated IRE1 clusters switch from growth to decay by finding the mean cluster size vs time for different $c_{\text{ext}}$ (Fig.~\ref{fig:Threshold}A). The threshold concentration between cluster growth and decay varies with tube width (Fig.~\ref{fig:Threshold}B), with the threshold suddenly increasing when the tube width increases above a particular diameter. The relatively low threshold concentrations for tube widths below this sudden increase in Fig.~\ref{fig:Threshold}B correspond to wrapped clusters, and the relatively high threshold concentrations for tube widths above this sudden increase correspond to round clusters. The tube diameter at which the threshold concentration changes suddenly increases approximately corresponds to the diameter at which clusters transition between wrapped and round conformations (Fig.~\ref{fig:ClosedSystem}C). Below and above this diameter at which the threshold concentration suddenly changes (at which clusters switch from round to wrapped), the threshold concentration changes less with tube diameter.

For an IRE1 cluster at the threshold concentration $c_0$, the inward flux of proteins from outside the tube region under consideration to the cluster is proportional to the external concentration $c_{\text{ext}} = c_0$, the tube radius $r_{\text{tube}}$, and the probability of encountering the cluster at its axial position $p_{\text{e}}$, and inversely proportional to the length of the tube $\ell$ (proportional to the probability of a protein at one end of the tube reaching the other end without first leaving by the initial end).
The inward flux is $\Phi_{\text{in}} = \alpha c_0 r_{\text{tube}}p_{\text{e}}/\ell$, with $\alpha$ a proportionality constant that includes factors other than concentration and geometry, such as the IRE1 diffusivity and protein interaction strength. For wrapped clusters, $p_{\text{e}}^{\text{wrap}} = 1$, because the cluster wraps around the tube, such that proteins cannot reach the axial position of the cluster without encountering the cluster. For round clusters, $p_{\text{e}} = 2r_{\text{cluster}}/(2\pi r_{\text{tube}}) = r_{\text{cluster}}/(\pi r_{\text{tube}})$ as a round cluster only covers a fraction of the tube circumference.

The flux of proteins escaping from the cluster will be proportional to the cluster interface length: $\Phi_{\text{out}}^{\text{wrap}} = \beta_{\text{wrap}}L_{\text{wrap}}$ and $\Phi_{\text{out}}^{\text{round}} = \beta_{\text{round}}L_{\text{round}}$, with $\beta$ a proportionality constant including IRE1 diffusivity, interaction strength, and the interface curvature (curvature will differ between wrapped and round conformations). $L_{\text{wrap}} = 4\pi r_{\text{tube}}$ and $L_{\text{round}} = 2\pi r_{\text{cluster}}$.

At the threshold concentration $c_0$, the cluster is not growing or decaying, and $\Phi_{\text{in}} = \Phi_{\text{out}}$. This yields $c_0^{\text{wrap}} = 4\pi\beta_{\text{wrap}}\ell/\alpha$ and $c_0^{\text{round}} = 2\pi^2\beta_{\text{round}}\ell/\alpha$. The difference between $\beta_{\text{wrap}}$ and $\beta_{\text{round}}$ is that the round clusters are expected to have a more curved interface than wrapped clusters. However, this curvature difference is expected to become small for larger clusters, such that $\beta_{\text{wrap}} \approx \beta_{\text{round}}$. With this approximation, the ratio between threshold concentrations is
\begin{equation}
\label{eq:ratio}
\frac{c_0^{\text{round}}}{c_0^{\text{wrapped}}} \approx \frac{\pi}{2}\ ,
\end{equation}
approximately 1.57. In Fig.~\ref{fig:Threshold}B, this ratio is approximately 1.64 for clusters with 100 proteins and 1.48 for clusters with 200 proteins (comparing the threshold concentrations at the smallest and largest tube diameters in Fig.~\ref{fig:Threshold}B). Round clusters on wider tubes require a higher external IRE1 concentration to avoid cluster decay compared to wrapped clusters on narrow tubes. This suggests that tube geometry (diameter) could play an important role in IRE1 cluster stability, and impact the location of persistent clusters as cluster coarsening occurs via Ostwald ripening. Figure~\ref{fig:Threshold} suggests that stability of clusters of a certain size is step-like, with clusters above a certain size gaining considerable stability on sufficiently narrow tubes.\\

\begin{figure}[t] 
    \begin{center}
    \includegraphics[width = 0.5\textwidth]{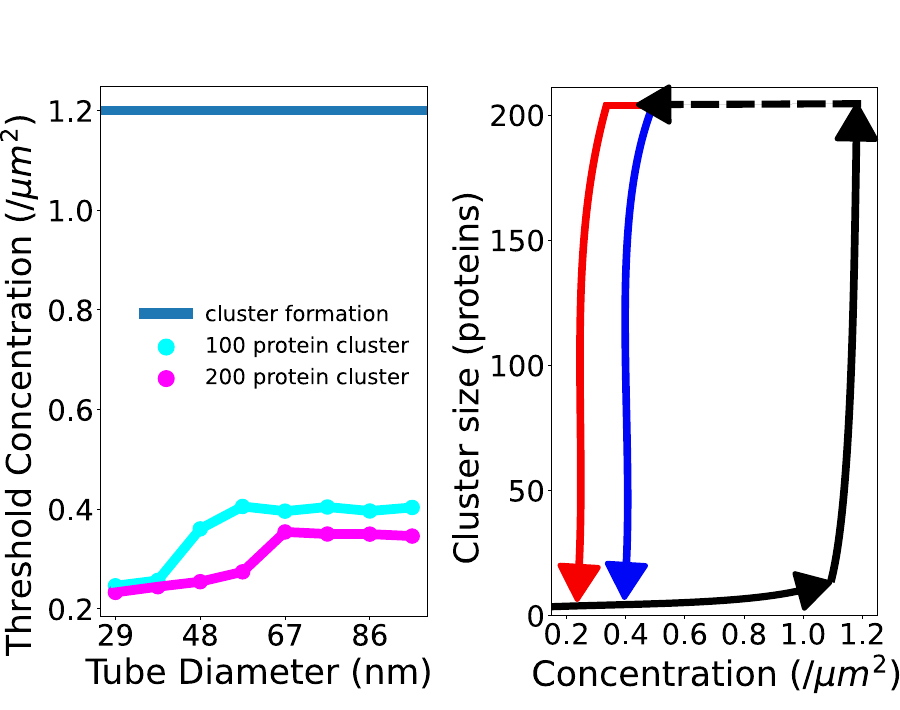}
    \caption{Summary of IRE1 concentrations important for cluster dynamics.
    Left: Threshold concentrations for cluster formation (blue, see Fig.~\ref{fig:OpenPD}), decay of 100- and 200-protein clusters (cyan and magenta, respectively, from Fig.~\ref{fig:Threshold}B). Clusters do not form until a relatively high IRE1 concentration is reached --- as concentration decreases, clusters can grow below this high cluster formation threshold. Clusters will decay on wider tubes at a higher concentration than on narrower tubes, because clusters on narrower tubes are in a wrapped configuration and are more stable, compared to round clusters on wider tubes.
    Right: Schematic of cluster growth and decay trajectory. Concentration begins near zero, with no cluster. As the concentration increases past the cluster formation threshold, clusters will form and grow (black solid curve). As the concentration begins to decrease, the cluster growth diminishes (dashed black curve). Once the concentration is sufficiently low, first round clusters (blue solid curve) and then wrapped clusters (red solid curve) will decay.
    $J=5.3\text{ }k_{\text{B}}T$ for both panels.}
    \label{fig:pd_and_hyst}
    \end{center}
\end{figure}

Overall, our results indicate hysteretic behavior for IRE1 cluster dynamics on ER tubes (see Fig.~\ref{fig:pd_and_hyst}). Under ER protein stress, the concentration of activated IRE1 will rise. Once the concentration has sufficiently risen (to approximately $1/\mu\text{m}^2$ for the parameters of Fig.~\ref{fig:pd_and_hyst}, comparable to physiological IRE1 concentrations of approximately $1/\mu\text{m}^2$~\cite{belyy2020quantitative,aragon2009messenger,west20113d}), IRE1 clusters can form. The available concentration of IRE1 in the ER network will decrease as IRE1 proteins join clusters and the clusters undergo Ostwald ripening. With time, as the actions stimulated by the unfolded protein response (UPR) pathway begin to mitigate ER protein stress, IRE1 may begin to deactivate, also reducing the pool of active IRE1 available for clustering. As the concentration of active IRE1 outside of clusters decreases, this concentration will first fall below the threshold concentration between cluster growth and decay for round clusters ($\sim 0.4/\mu\text{m}^2$ in Fig.~\ref{fig:pd_and_hyst}), which is higher than this threshold concentration for wrapped clusters ($\sim 0.25/\mu\text{m}^2$ in Fig.~\ref{fig:pd_and_hyst}) by approximately a factor $\pi/2$; and then fall below the threshold concentration between wrapped cluster growth and decay. Wrapped clusters will be the last to begin to decay as the active IRE1 concentration decreases. There is a gap between the threshold concentration between cluster growth and decay (low) and the concentration required for cluster formation (high). As concentration of active IRE1 proteins is rising from nearly zero, the concentration must become relatively high to cause cluster formation. Once formed, clusters will continue to grow at concentrations below the cluster formation threshold.

\section{Discussion}

IRE1 protein clusters form on the surface of the endoplasmic reticulum (ER) under conditions of unfolded protein stress in the ER. As the ER is a network of tubes and sheets, many of these IRE1 clusters form on the tube surfaces. The tubular geometry allows a growing cluster to `wrap' around the tube and grow without further increases to the cluster interface length. Using kinetic Monte Carlo simulations, we have shown that tube diameter affects cluster dynamics, as narrower tubes encourage cluster wrapping, affecting cluster growth and stability.

The cluster growth rate increases when the cluster is sufficiently large that the cluster has transitioned from a round to a wrapped conformation. When clusters are wrapped around the tube, they gain several advantages for growth relative to round clusters. Wrapped clusters have an interface that does not grow with increased cluster size, in contrast to round clusters which have an interface length that increases with the square root of the cluster area, and protein escape from clusters will increase with a longer interface. Wrapped clusters have a flat interface, while round clusters have a curved interface --- lower interface curvature decreases protein escape from clusters~\cite{krishnamachari1996gibbs}. Wrapped clusters are also able to encounter all proteins traversing the tube, in contrast to round clusters. However, clusters on wider tubes grow faster than those on narrower tubes, although clusters transition from round to wrapped at a smaller size on narrower tubes, due to the higher rate of proteins entering wider tubes exposed to the same external concentration, but with greater circumference, compared to narrower tubes. Overall we find that tube geometry is important to determining cluster growth rate, both because narrow tubes allow a wrapped cluster conformation that provides a growth increase relative to a round conformation, and because clusters on wider tubes grow more quickly.

We find that for clusters of a given size, wrapped clusters on narrower tubes decay significantly more slowly than round clusters on wider tubes. By examining cluster decay under conditions with no incoming proteins to grow the cluster, we isolated the impact of protein escape from clusters, showing that proteins escape from wrapped clusters more slowly than from round clusters. We attribute this slower escape from wrapped clusters to a shorter interface length and less curved interface compared to round clusters.

We find that the threshold IRE1 concentration for IRE1 cluster formation does not depend on tube diameter. However, the growth and decay of existing clusters depends on tube diameter, particularly through the control of cluster conformation (wrapped vs round) by the tube diameter. We show that wrapped clusters on narrower tubes are more stable than round clusters on wider tubes, by demonstrating that the threshold protein concentration between cluster growth and decay is higher for the round clusters on wide tubes than for equal-sized wrapped clusters on narrow tubes. This lower threshold concentration between growth and decay for wrapped clusters is attributed to less protein escape due to a shorter and less curved interface, and to the encounter of all proteins traversing the tube with the cluster. We predict that wrapped clusters will switch from growth to decay at protein concentrations that are lower by approximately a factor $\pi/2$, which is similar to our simulation results.

We also find that the threshold concentrations between growth and decay of existing clusters (for both round and wrapped clusters) are substantially lower than the cluster formation threshold. Cluster dynamics are thus hysteretic. Clusters do not form until a high concentration is reached, but once the clusters have formed, they can grow at concentrations substantially below the cluster formation concentration threshold. Cluster growth will decrease the concentration of available IRE1, evaporating all but the largest clusters; and the unfolded protein response (UPR) will also act to reduce the active IRE1 concentration. The cluster dynamics thus have a form of `memory', because the cluster formation concentration threshold is higher than the threshold between the growth and decay of existing clusters.

It is known that Ostwald ripening will lead small clusters to decay prior to larger clusters~\cite{yao1993theory}. Our results suggest that cluster wrapping adds an additional effect, where larger IRE1 clusters are able to wrap around the ER tube and decrease the threshold concentration between cluster growth and decay. For wider ER tubes, clusters must be larger to achieve a wrapped conformation.  As the cluster numbers decrease through Ostwald ripening, our results suggest that wrapped clusters, perhaps more likely to be found on narrow ER tubes, will be among the clusters that grow most persistently and will be among the last to begin to decay. The slower decay dynamics, at lower IRE1 concentration, of IRE1 clusters wrapped on narrow ER tubes thus control the persistence of IRE1 clustering and downstream signaling.

Our simulation results showing that sufficiently large clusters prefer to wrap around the ER tube align with experimental observations of IRE1 clusters. IRE1 clusters have been in observed with diverse morphologies, including those that appear wrapped around ER tubes~\cite{belyy2020quantitative,tran2021stress} and localize to narrow ER tubes~\cite{tran2021stress}.


We have shown that domain geometry can affect cluster behavior, with cluster wrapping around tubes permitting narrower tubes to harbor more stable clusters. Previous work investigating autophagy receptor protein cluster formation on peroxisomes has shown that clusters are more likely to form and grow on larger spheres~\cite{brown2015cluster,brown2017model}. While both represent local geometric influence over cluster behavior, clusters on tubes enhance cluster stability by altering the cluster interface compared to round clusters, while cluster growth on larger spheres is mediated by initial formation of larger clusters with an Ostwald ripening growth advantage and the arrival of more proteins to larger spheres. Clusters forming on a finite two-dimensional surface have a minimum stable cluster size that increases as the finite two-dimensional area grows~\cite{krishnamachari1996gibbs}, similar to our finding that a smaller domain can increase the stability of smaller clusters.

Yeast ER tubes (mean diameter of 38 nm~\cite{west20113d}) are significantly narrower than mammalian ER tubes (mean diameter of 96 nm~\cite{schroeder2019dynamic}), suggesting IRE1 cluster wrapping could occur for smaller clusters and be more stable in yeast compared to mammalian cells.
Recent experiments suggest that mammalian ER tubes have narrow ($\sim$25 nm diameter) and wide ($\sim$100 nm diameter) regions along their length, and that ER tubes in different cell types may be primarily the narrow or wide type~\cite{wang2022endoplasmic}. Tube radii in both yeast (20 -- 50 nm diameter~\cite{west20113d}) and mammalian (50 -- 140 nm diameter~\cite{schroeder2019dynamic}) cells also vary widely. These observations, in combination with our results, suggest that IRE1 clusters could localize to or be more stable on narrow ER tube regions or narrower ER tubes. Clusters located in narrow regions or on narrow tubes would experience an energy difference if moved from a region where the cluster could wrap to where the cluster must be round (in addition to the energy barrier of cluster conformation change). It also suggests that cell types with largely narrow tubes could be more conducive to IRE1 cluster stability and persistence.

Many of the experiments that observe IRE1 clustering overexpressed IRE1. Despite work indicating that IRE1 clusters are important for UPR signaling~\cite{korennykh2009unfolded,li2010mammalian,belyy2020quantitative,tran2021stress}, recent work~\cite{belyy2022endoplasmic,gomez2022live} indicates that endogenous IRE1 levels in certain mammalian cells (approximately $1/\mu\text{m}^2$) are significantly lower than when IRE1 is overexpressed, 
and that large IRE1 clusters may not be required for UPR signaling. However, it is noted that IRE1 concentration can vary between cells and for cells in pathological states~\cite{belyy2022endoplasmic,harnoss2019disruption,harnoss2020ire1alpha}. For yeast, experiments demonstrate that IRE1 cluster formation upon ER stress in yeast is essential for UPR signaling~\cite{kimata2007two,aragon2009messenger,van2014specificity}  --- we are not aware of any work suggesting otherwise. If there is a difference in IRE1 clustering behavior between yeast and mammalian cells, it is possible that the smaller ER tubes in yeast cells ($\sim$40 nm diameter in yeast~\cite{west20113d} vs $\sim$100 nm diameter in mammals~\cite{schroeder2019dynamic}) combined with the increased cluster stability provided by cluster wrapping on narrower tubes could contribute to more IRE1 clustering behavior in yeast.  Although large IRE1 clusters may not be required for UPR signaling to occur~\cite{belyy2022endoplasmic,gomez2022live}, there are IRE1 signaling modes that require larger IRE1 oligomers~\cite{le2021decoding}, and our work explores how the persistence of these signaling modes will be affected by IRE1 clustering and ER geometry.

IRE1 clustering on the ER may be tied to human health.
Prolonged UPR activation (including IRE1 signaling) is associated with neurodegenerative diseases, suggesting a possible pathological role for dysfunctional IRE1 and other UPR signaling~\cite{scheper2015unfolded}. Insulin production in pancreatic beta cells induces ER stress and UPR (including IRE1) activation, and it has been proposed that UPR dysregulation may predispose individuals to diabetes~\cite{shrestha2021pathological}.
Protein clustering on other organelles, such as mitochondria, suggests that further understanding of intracellular receptor clustering dynamics may be important for human health. Mitochondrial MAVS proteins aggregate as part of viral immunity. Typically dispersed on the membranes of many mitochondria, MAVS proteins aggregate on the membranes of a few mitochondria as part of antiviral innate immune response when a cell detects viral RNA, and are required for RLR viral infection signaling~\cite{onoguchi2010virus,hou2011mavs}. Persistent MAVS aggregation may play a role in lupus pathology~\cite{shao2016prion}.

With Monte Carlo simulations, we have found that narrow ER tubes facilitate an IRE1 cluster conformation that wraps around the tube, influencing cluster growth and increasing cluster stability and persistence. As IRE1 is an important protein in signaling of the unfolded protein response, our work shows that geometry can be an important factor for modulating cell signaling and maintaining cell health.

\acknowledgements
This work was supported by a Natural Sciences and
Engineering Research Council of Canada (NSERC) Discovery Grant (A.I.B.) and by start-up funds provided by the Toronto Metropolitan University Faculty of Science (A.I.B.), and was enabled
by computational resources provided by the Digital Research Alliance of Canada (alliancecan.ca), including a cluster usage resource allocation (A.I.B.).
The authors thank Elena Koslover (UC San Diego), Sean Cornelius, and Eric De Giuli (Toronto Metropolitan University) for useful discussions and feedback.

\appendix

\section{Method} \label{method.tex}

To simulate IRE1 cluster dynamics on ER tubes we used the widely-applied kinetic Monte Carlo (kMC) algorith~\cite{andersen2019practical,bortz1975new,andersen2019practical,cheimarios2021monte} with IRE1 proteins diffusing on a two-dimensional lattice.

The diffusivity on a two-dimensional lattice with spacing $\Delta x$ is $D = \Delta x^2 / (4\Delta \tau)$, with $\tau$ the mean time between subsequent steps. The IRE1 diffusivity on the ER membrane has been measured as $D = 0.24\text{ }\mu \text{m}^2/\text{s}$~\cite{belyy2020quantitative}. For the lattice spacing $\Delta x$ we will use the diameter of an IRE1 protein, which is approximately 10 nm~\cite{tran2021stress}. This sets $\tau \approx 10^{-4}\text{ s}$.

While ER tube lengths have significant variation, many tube lengths fall in the range of 0.5 -- $3\text{ }\mu \text{m}$ in mammalian cells~\cite{georgiades2017flexibility} and 250 -- 750 nm in yeast~\cite{west20113d}, and we use a tube length of 1 $\mu\text{m}$ for all simulations. ER tube diameter has a wide range, in mammals largely falling from 50 -- 140 (mean 96 nm)~\cite{schroeder2019dynamic} with narrower examples with diameters as small as 25 nm observed~\cite{tran2021stress,wang2022endoplasmic}, and in yeast largely from 20 -- 50 nm (mean 38 nm)~\cite{west20113d}. In our simulations, we explore diameters from 28 nm to 121 nm.

In the direction of the axis perpendicular to the length of the tube, a diffusing protein has periodic boundary conditions. While the results of Fig.~\ref{fig:ClosedSystem} are for a closed tube without IRE1 protein exchange beyond the tube, all other results allow proteins to diffuse into and out of the tube ends. IRE1 proteins enter lattice sites at tube ends subject to a constant IRE1 protein concentration $c_{\text{ext}}$ boundary condition. With $c_{\text{ext}}$ expressed in units of proteins/$\mu$m$^2$, and a concentration of 1/$\mu$m$^2$ = 10$^{-4}$/lattice site, proteins enter empty edge lattice sites at a rate $10^{-4}\text{}\mu\text{m}^2c_{\text{ext}}/(4\tau)$.

The IRE1 copy number in a mammalian cell has been estimated at $10^4$, and the ER surface area estimated as $10\times$ the cell membrane area or $10^4\text{ }\mu\text{m}^2$, giving an IRE1 concentration of 1/$\mu$m$^2$~\cite{belyy2020quantitative}. This corresponds to $\sim$0.31 IRE1 proteins on a tube with a 100 nm diameter and 1 $\mu$m length. The IRE1 copy number in a yeast cell has been estimated at 250~\cite{aragon2009messenger}. The ER volume in a 1 $\mu$m diameter bud has been estimated at 0.025 $\mu$m$^3$~\cite{west20113d}, and an approximately spherical yeast cell has volume $42\text{ }\mu\text{m}^3$~\cite{brown2022mitochondrial} corresponding to a 4.32 $\mu$m diameter, so scaling the ER volume by the cell/bud volume ratio, the ER volume in the mother cell is approximately 2 $\mu$m$^3$. The ratio of volume to surface area in yeast ER is approximately 8 nm~\cite{west20113d}, so the yeast ER surface area is approximately 250 $\mu$m$^2$. Thus the approximate IRE1 concentration on the ER surface is 1/$\mu$m$^2$, very similar to the mammalian concentration. We explore IRE1 concentrations of zero to see how clusters will decay; within an order of magnitude of the physiological concentration $1/\mu\text{m}^2$; and up to $34/\mu\text{m}^2$ to encourage fast cluster growth.

IRE1 interaction strengths $J$ in the range 3 -- 7 $k_{\text{B}}T$ were used. $J = 5$ -- $7\text{ }k_{\text{B}}T$ lead to cluster formation near the cellular IRE1 concentration and $J = 3$ -- $4\text{ }k_{\text{B}}T$ provide faster dynamics for efficient simulations.

The simulation allows IRE1 proteins forming a cluster to collectively diffuse, as IRE1 clusters freely diffuse on the ER membrane, rather than experiencing constrained diffusion or active transport~\cite{belyy2020quantitative}. Following experiments showing diffusivity in a membrane scaling with inverse radius~\cite{gambin2006lateral} and consistent with the Stokes-Einstein relation~\cite{luo2015kinetic}, cluster diffusivity is scaled as $1/R$ where $R$ is the cluster radius or $N^{-1/2}$, where $N$ is the number of proteins in the cluster. In the simulation, IRE1 proteins can join clusters as multimers, but only escape as individual proteins. Clusters are also not permitted to leave the tube on the ends, with such Monte Carlo steps not allowed.


A cluster is considered wrapped if it has at least one protein in every row of the lattice. In Fig.~\ref{fig:ClosedSystem}B, the cluster conformation transition frequency is calculated by simulating for a fixed time, and then dividing the number of transitions by the simulation time. One transition is counted for each switch from round to wrapped conformation or vice versa.
In Fig.~\ref{fig:OpenSystem}A the round-to-wrapped transition time is the average of the first time the cluster transitions from round to wrapped and the final time the cluster transitions from round to wrapped, as the cluster can fluctuate between the two conformations.
In Fig.~\ref{fig:Decay}B, the cluster decay rate is the slope between the first and last data point in a decay curve averaged over 30 simulations.

\section{Additional Simulations}

\begin{figure*} 
\includegraphics[width = 0.5\textwidth]{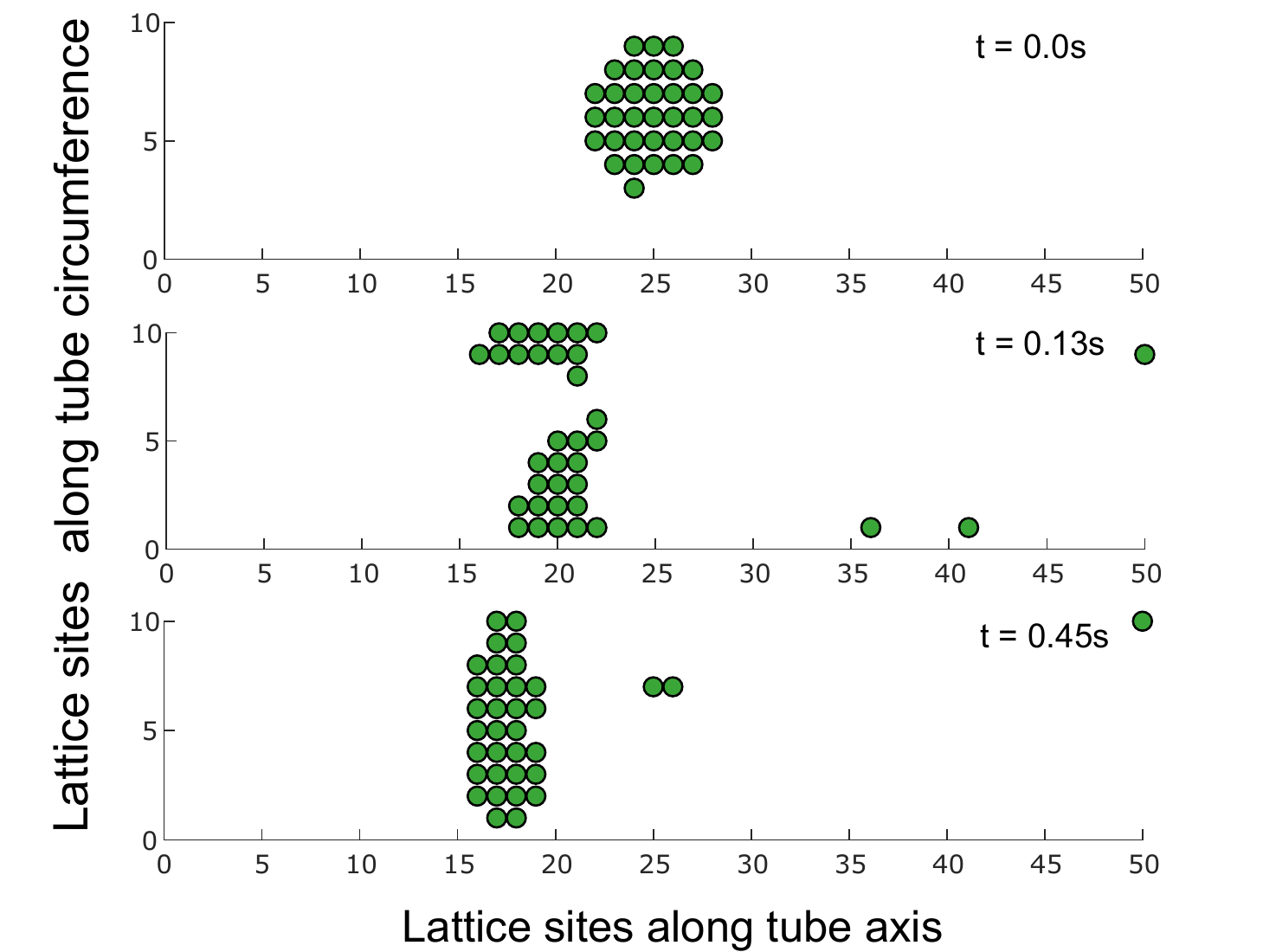}
\caption{Snapshots of cluster wrapping transition. The lattice has periodic boundary conditions in the direction long the tube circumference. (A) Cluster is in an approximately round state. (B) Cluster is growing long in the direction of the circumference o the tube. (C) The cluster has wrapped around the tube. Closed boundary conditions at tube edges, $J = 3 k_BT$, and 35 proteins in approximately 32 nm diameter and length 500 nm tube.
}
\label{fig:cluster_snapshots}
\end{figure*}

Figure~\ref{fig:cluster_snapshots}A,B,C shows the transition from a round cluster to a wrapped cluster.

\begin{figure*} 
\includegraphics[width = \textwidth]{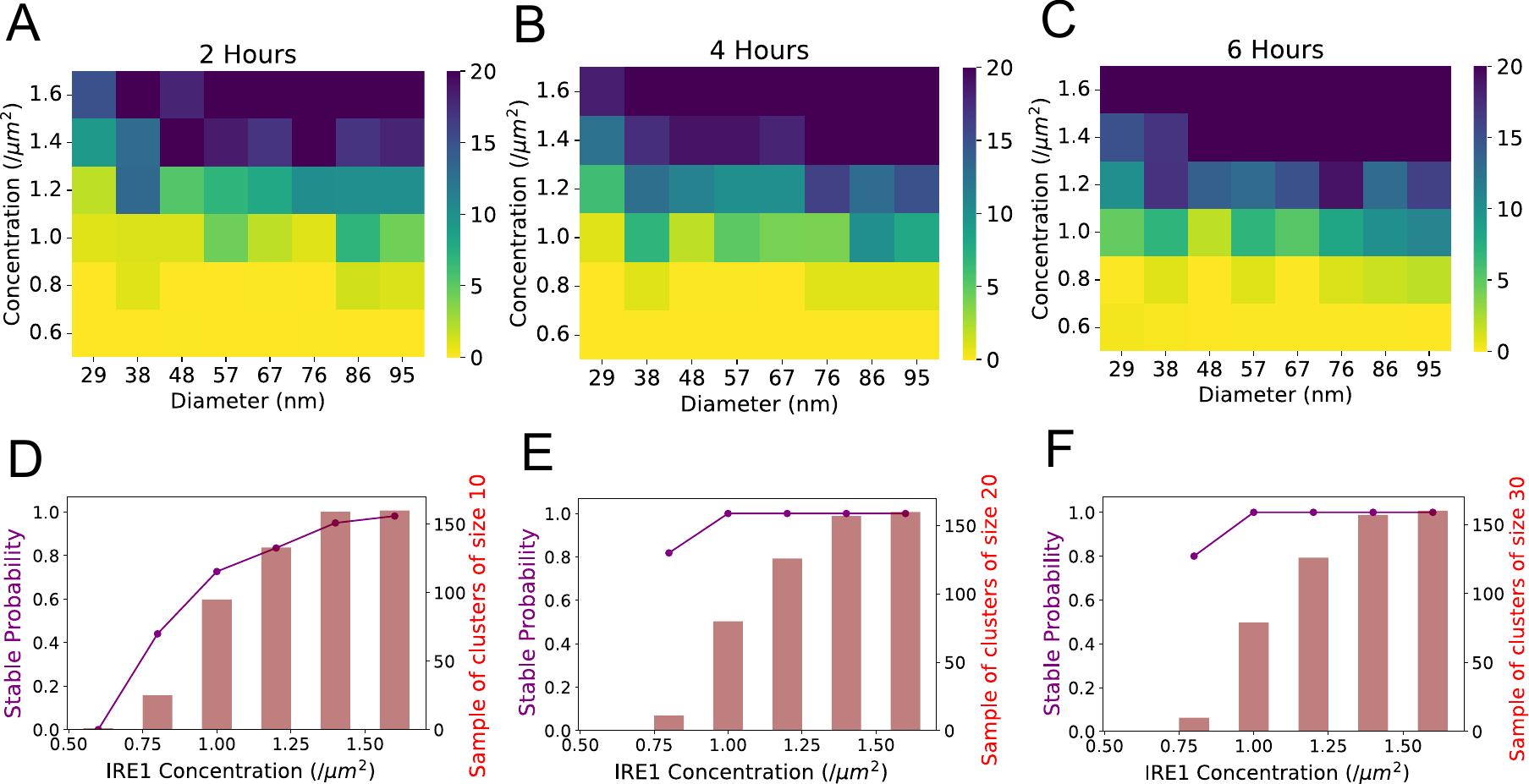}
\caption{ Cluster formation with open boundary conditions. (A,B,C) mean IRE1 cluster size  with open boundary conditions at the indicated times. The IRE1 external (to the tube region under consideration) concentration $c_{\text{ext}}$ and the tube diameter are varied,
with color map indicating mean number of proteins in a cluster. Clusters of sizes greater than 20 are binned with cluster size 20. (D,E,F) Cluster decay probability for clusters of different size with different external IRE1 concentrations $c_{\text{ext}}$. The left axis (magenta) is the probability that a cluster of size 10 (D), 20 (E), and 30 (F) will decay. The right axis is the number of clusters included in each decay probability calculation. Tube length 1 $\mu$m and $J = 5.3\text{ }k_{\text{B}}T$.
}
\label{fig:phase_d_details}
\end{figure*}

Figure~\ref{fig:phase_d_details}A,B,C shows cluster formation with open boundary conditions for different times, after 2 hours of cluster development from a tube with a single protein (Fig.~\ref{fig:phase_d_details}A), 4 hours (Fig.~\ref{fig:phase_d_details}B), and 6 hours (Fig.~\ref{fig:phase_d_details}A, which is shown in Fig.~\ref{fig:OpenPD}). Cluster formation, shown for various concentrations and tube diameters, does not appear to meaningfully change across these times 2 -- 6 hours.

Figure~\ref{fig:phase_d_details}D,E,F shows the probability of cluster decay for clusters having reached 10 proteins (Fig.~\ref{fig:phase_d_details}D), 20 proteins (Fig.~\ref{fig:phase_d_details}E), and 30 proteins (Fig.~\ref{fig:phase_d_details}F). A cluster of size 10 is likely to decay, particularly at lower concentrations. However, clusters that have reached size 20 or 30 are very unlikely to decay, even at lower concentrations. This allows us to consider that clusters that have reach a size of 20 are unlikely to decay and the tube will continue to harbor a cluster.

\begin{figure*} 
\includegraphics[width = 0.8\textwidth]{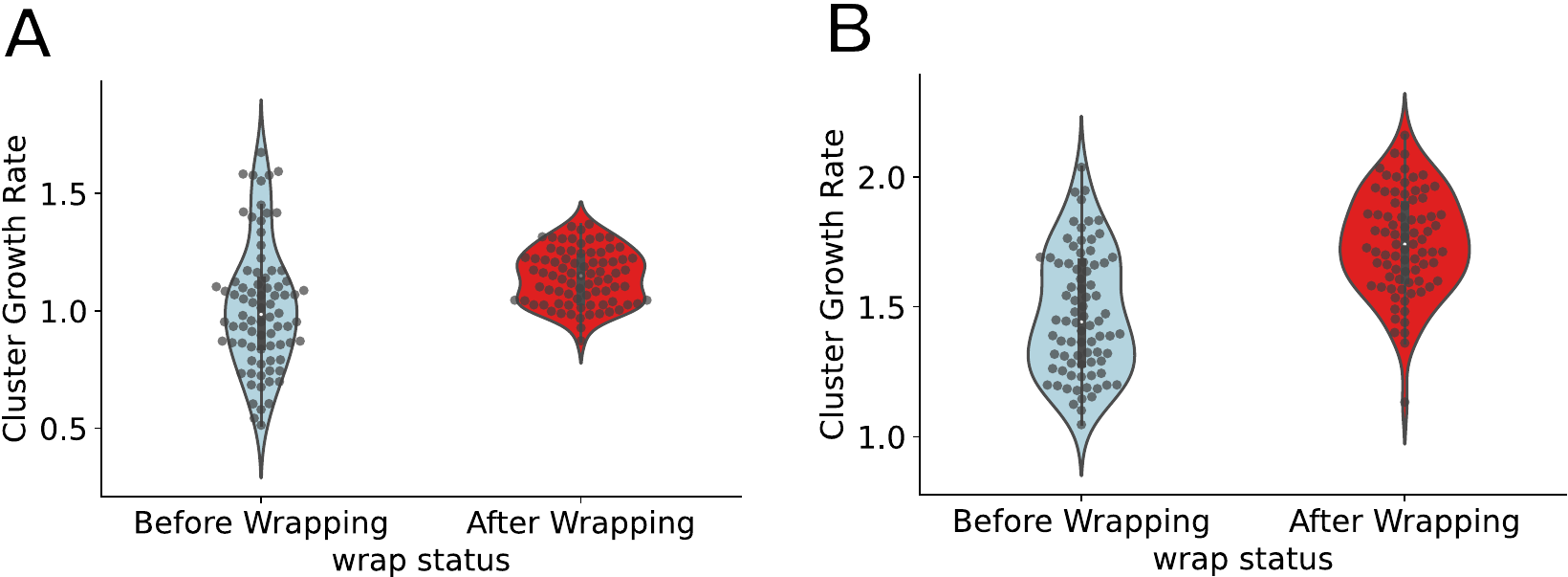}
\caption{Violin plot
of the cluster growth rates before and after the cluster wrapping transition. (A) Tube diameter of 32 nm. Mean growth rate before wrapping is 1.01 proteins/s, and after wrapping it is 1.14 proteins/s. The Kolmogorov-Smirnov (K-S)  test returned a p-value of $4.27\times 10^{-8}$ between the two distributions.  (B) Tube diameter of 46 nm., Mean growth rate before wrapping is 1.47 protein/s, and after wrapping it is 1.75 proteins/s. The K-S test returned a p-value of $6.20\times 10^{-10}$ between the two distributions. For both panels, $J = 5.3\text{ } k_{\text{B}}T$ as in Fig.~\ref{fig:OpenPD}, tube length is $1 \mu \text{m}$, $c_{\text{ext}} = 1 / \mu \text{m}^2$, and means averaged over 80 runs.}
\label{fig:Slope_change_additional}
\end{figure*}

Figure~\ref{fig:Slope_change_additional}A,B shows the change in cluster growth rates before and after a wrapping transition, as in Fig.~\ref{fig:OpenSystem}B. Cluster growth rate increases after the wrapping transition.

\section{Derivation}
In the derivation of Eq.~\ref{eq:ratio} there is an inverse length $\ell$ of the tube factor for the inward flux $\Phi_{\text{in}}$. Here we consider a protein taking steps on a discrete lattice away from an absorbing boundary, to represent the probability that a protein that has just entered the tube region under consideration (and is inside by a distance equal to a single lattice spacing) will reach a distance $\ell$ into the tube region under consideration without first reaching the absorbing boundary.

The probability that a protein a single lattice site into the tube will move to the second lattice site into the tube without hitting the absorbing boundary is 1/2. More generally, $P_{n-1\to n,\text { w/o}\to0}$ is the probability that a protein $n-1$ sites from the absorbing boundary will reach site $n$ without first reaching the absorbing boundary at site 0,
\begin{align}
P_{n-1\to n,\text { w/o}\to0} &= \frac{1}{2} + \frac{1}{2} P_{n-2\to n-1,\text { w/o}\to0} \frac{1}{2} \nonumber\\
&+  \left(\frac{1}{2} P_{n-2\to n-1,\text { w/o}\to0}\right)^2\frac{1}{2} + \dots \label{eq:derivfirst}\\
&= \frac{1}{2}\sum_{m=0}^{\infty} \left(\frac{1}{2}P_{n-2\to n-1,\text { w/o}\to0}\right)^m \\
&= \frac{1}{2} \left(1 - \frac{1}{2}P_{n-2\to n-1,\text { w/o}\to0}\right)^{-1}\label{eq:derivsecond} \ .
\end{align}
The first term in Eq.~\ref{eq:derivfirst} represents a protein that immediately steps from site $n-1$ to site $n$ without other intervening steps. The second term in Eq.~\ref{eq:derivfirst} represents a protein that first steps from site $n-1$ to site $n-2$ (probability 1/2) and then follows a trajectory that returns to site $n-1$ without hitting the absorbing boundary (probability $P_{n-2\to n-1,\text { w/o}\to0}$) and then takes a step to site $n$ (probability 1/2). Subsequent terms involve $m$ of these steps to site $n-2$, and then trajectories that return to site $n-1$ without hitting the absorbing boundary.

Equation~\ref{eq:derivsecond} is a sequential formula for $P_{n-1\to n,\text { w/o}\to0}$. We can start with $P_{1\to 2,\text { w/o}\to0} = 1/2$ and find $P_{2\to 3,\text { w/o}\to0} = 2/3$, and similarly that $P_{2\to 3,\text { w/o}\to0} = 3/4$. Generally, for $P_{n-2\to n-1,\text { w/o}\to0}=(n-2)/(n-1)$ then
\begin{align}
P_{n-1\to n,\text { w/o}\to0} &= \frac{1}{2} \left(1 - \frac{1}{2}\frac{n-2}{n-1}\right)^{-1}\\
&=\frac{n-1}{n} \ .
\end{align}
The probability of reaching a site $n$ lattice sites from the absorbing boundary is then
\begin{align}
    P_{1\to n,\text { w/o}\to0} &= \prod_m^{n} \frac{n-1}{n} \\
    &= \frac{1}{n} \ .
\end{align}
The number of sites $n = \ell/\Delta x$, so the probability of reaching a distance $\ell$ into the tube is $\Delta x/\ell$. In the derivation of Eq.~\ref{eq:ratio}, the $\Delta x$ factor is absorbed into $\alpha$.


\providecommand{\noopsort}[1]{}\providecommand{\singleletter}[1]{#1}%

\end{document}